\documentclass[letterpaper,12pt]{article}

\usepackage{amsmath,amssymb,amsfonts,graphicx,subfigure,cite,url}

\title{\bf \Large Finding the Center of Mass of a Soft Spring}
\author{Juan D. Serna\thanks{serna@uamont.edu} \\
        School of Mathematical and Natural Sciences \\
        University of Arkansas at Monticello, Monticello, AR 71656 \\
\and
        Amitabh Joshi\thanks{ajoshi@eiu.edu} \\
        Department of Physics \\
        Eastern Illinois University, Charleston, IL 61920}
\date{November 13, 2011}

\begin{document}
\maketitle

\begin{abstract}
This article shows how to use calculus to find the center of mass of a soft,
vertically suspended, cylindrical helical spring, which necessarily is
non-uniformly stretched by the action of gravity. A general expression for the
vertical position of the center of mass is obtained and compared with other
results in the literature.
\end{abstract}

\newcommand*{\half}{\frac{1}{2}}

\section{Introduction}
Finding the center of mass is a fundamental part of the mechanics of rigid
bodies. The center of mass of highly symmetric rigid bodies is easily obtained.
However, if the shape and size of a body is affected by external factors, like
gravity, its center of mass changes in a way that it is sometimes difficult to
predict. This is the case with a soft helical spring. When such a spring is hung
vertically it stretches non-uniformly, and its center of mass is no longer
located at the center of the spring's length.

In this article, we find the center of mass of a vertically suspended soft
spring (a Slinky) using the calculus. Here, we use the term ``soft'' to describe
a spring whose stiffness is small enough so that its own weight stretches it
noticeably when it is suspended vertically. Our results are in excellent
agreement with experimental results and non-analytic approaches. The method we
use is easily incorporated into any undergraduate calculus or general physics
course.

\section{Vertical elongation}
Consider a cylindrical helical spring of total mass $m_{\mathrm{s}}$, relaxed
length $L_0$ (neither stretched nor compressed), and stiffness constant $k$. If
the spring is suspended vertically, the mass per unit length becomes a function
of position, and the spring stretches \textit{non-uniformly} to a new length
$L_0 + \Delta L$, where $\Delta L$ is the elongation due to the spring's own
weight. The static and dynamic properties of such a spring have been extensively
discussed by several authors, and expressions for the vertical position of the
$n$th turn and the stretch per unit mass have been
derived~\cite{Heard:1102,Galloni:1076,Lancaster:217,
Mak:994,French:244,Christensen:818}.

Our goal is to describe the spring elongation due to gravity in terms of
position coordinates \textit{only}. A convenient set of equations are those
given for the 3D helix in Cartesian coordinates $x(\theta)=r\cos\theta,\;
y(\theta)=r\sin\theta,\; z(\theta)=c\,\theta + \Delta z,$ where $r$ is the
radius of the spring, $c = L_0 / 2 \pi N$ is the pitch (the width of one
complete turn), $N$ is the number of turns of the spring, $\theta$ is an angular
parameter ($0 \le \theta \le 2 \pi N$), and $\Delta z$ is the spring's
elongation due to its own weight.

\begin{figure}[h]
\begin{center}
\subfigure[Horizontal spring.]{
\includegraphics[scale=0.65]{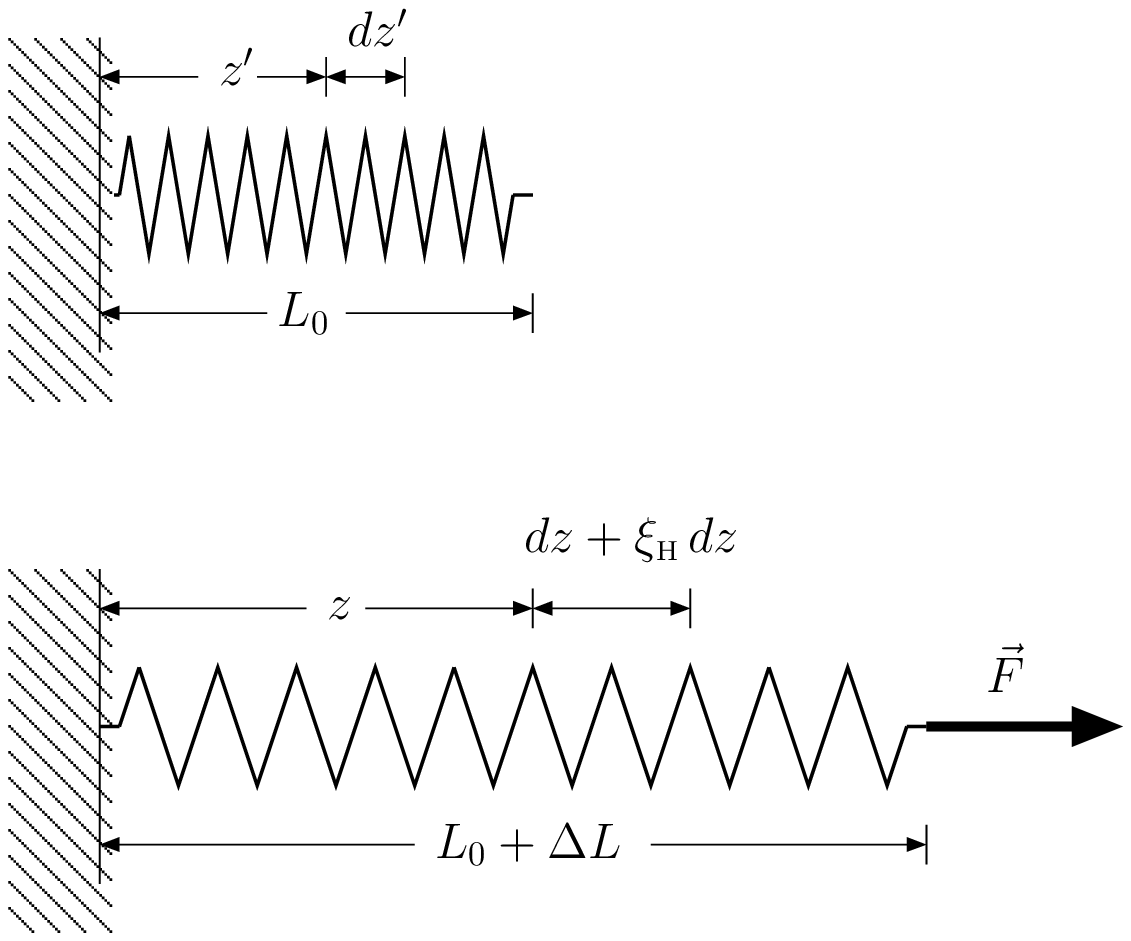}}
\hskip 1.0cm
\subfigure[Suspended spring.]{
\includegraphics[scale=0.65]{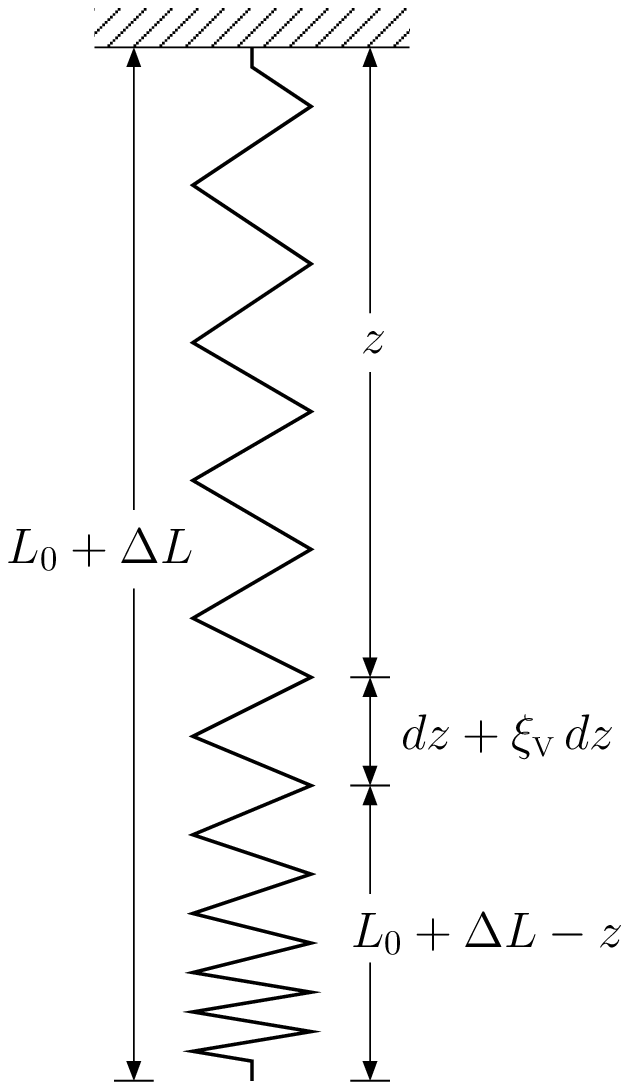}}
\caption{\label{Fig:hvSprings} Elongations $\xi_{\mbox{\scriptsize H}}$ and
$\xi_{\mbox{\scriptsize V}}$ for a segment $dz$ of a uniform spring stretched
(a) horizontally and (b) vertically.}
\end{center}
\end{figure}

To find $\Delta z$, we consider first, for simplicity, the elongation of a
horizontal spring pulled by a force $\vec{F}$ as shown in
Figure~\ref{Fig:hvSprings}(a). When the spring is not stretched, we use $z'$ as
the position variable to locate any particular turn. This variable runs from one
end of the spring up to its total \textit{natural length} ($0 \le z' \le L_0$).
Likewise, when the spring is stretched and gives rise to an elongation $\Delta
L$, we use the variable $z$ running from one end of the spring up to its maximum
length of elongation ($0 \le z \le L_0 + \Delta L$). From the geometry of the
problem,
\begin{align}\label{Eq:zProportion}
  \dfrac{z'}{z} = \dfrac{L_0}{L_0 + \Delta L}\,,
  \qquad \mbox{and} \qquad
  dz = \dfrac{L_0 + \Delta L}{L_0}\,dz'\,.
\end{align}

Let us consider a spring portion of length $dz$ located at a distance $z$ from
an end fixed to a wall. For a uniform spring, the elongation of this segment is
given by the direct proportion
\begin{align}\label{Eq:Proportion}
  \dfrac{\xi_{\mbox{\scriptsize H}}\,dz}{\Delta L} = \dfrac{dz}{L_0}\,,
\end{align}
with $\xi_{\mbox{\scriptsize H}}$ being the \textit{dimensionless} horizontal
elongation factor of the segment. Mak~\cite{Mak:994} showed that if $k'$ is the
stiffness constant of this segment, then from Hooke's law and
equation~\eqref{Eq:Proportion}, we have
\begin{align}\label{Eq:kSegment}
  k' dz = \dfrac{F}{\xi_{\mbox{\scriptsize H}}}
        = \dfrac{F}{(\Delta L/L_0)}
        = \dfrac{F}{\Delta L} L_0
        = k\,L_0\,,
\end{align}
with $k=F/\Delta L$ representing the stiffness constant of the whole spring.
(When a spring is divided into segments, each segment has a stiffness constant
different and bigger than the stiffness constant of the whole spring.)

On the other hand, if the spring is suspended vertically (see
Figure~\ref{Fig:hvSprings}(b)), then, accordingly to Galloni and
Kohen~\cite{Galloni:1076}, each segment $dz$ of the spring experiences an
elongation proportional to the weight of the spring below the segment,
$m_{\mathrm{s}}\,g\,(L_0 + \Delta L - z) / (L_0 + \Delta L)$, where $g$ is the
acceleration due to gravity. Following Hooke's law, we get
\begin{align}\label{Eq:vElongation}
  k'(\xi_{\mbox{\scriptsize V}}\,dz)
  = \dfrac{m_{\mathrm{s}}\,g\,(L_0 + \Delta L - z)}{L_0 + \Delta L}\,,
\end{align}
with $\xi_{\mbox{\scriptsize V}}$ the vertical elongation factor of the segment.
Using~\eqref{Eq:zProportion} and~\eqref{Eq:kSegment}, $k'$ and $dz$ can be
eliminated from~\eqref{Eq:vElongation}, and the elongation factor takes the form
\begin{align*}
\xi_{\mbox{\scriptsize V}}
= \dfrac{m_{\mathrm{s}}\,g}{k\,L_0}
\left( \dfrac{L_0 + \Delta L - \dfrac{L_0 + \Delta L}{L_0}\,z'}
                                     {L_0 + \Delta L} \right)
= \dfrac{m_{\mathrm{s}}\,g}{k\,L_0} \left( \dfrac{L_0 - z'}{L_0}\right)\,.
\end{align*}
This allows us to find the contribution $\Delta z$ to the vertical elongation of
the spring by integrating the elongation factor $\xi_{\mbox{\scriptsize V}}(z')$
from the top of the spring $z'=0$ to any particular turn $c\,\theta$, obtaining
\begin{align*}
\Delta z = \int_0^{c\,\theta} \xi_{\mbox{\scriptsize V}}(z')\,dz'
= \dfrac{m_{\mathrm{s}}\,g\,(2\,L_0 - c\,\theta)\,c\,\theta}{2\,k\,L_0^2}\,.
\end{align*}

\section{The center of mass}
To find the position of the center of mass of the suspended spring, we first
calculate its linear mass density $\lambda$. If we assume that the mass of the
spring is uniformly distributed along the entire helix, then the density takes
the form $\lambda= m_{\mathrm{s}}/S$, where $S$ is the total arc length of the
spring. Since the contribution $\Delta z$ to the vertical elongation of the
spring  does not affect the length $S$, the infinitesimal element of arc can be
written as $ds^2 = dx^2 + dy^2 + d(c\,\theta)^2$, and the total arc length of
the spring calculated as follows
\begin{align*}
S = \int_0^{2\pi N}\sqrt{\left[\dfrac{dx}{d\theta}\right]^2 +
\left[\dfrac{dy}{d\theta}\right]^2 +
\left[\dfrac{d(c\,\theta)}{d\theta}\right]^2}\,d\theta
  = \int_0^{2 \pi N} \sqrt{r^2 + c^2}\,d\theta
  = 2 \pi N \sqrt{r^2 + c^2}\,,
\end{align*}
where the lower and upper bounds of the integral represent the starting and
ending turns of the spring. Now, the position of the center of mass can be
determined from the equations
\begin{align}\label{Eq:CM}
X_{\mathrm{cm}} = \dfrac{\int x\,dm}{m_{\mathrm{s}}}\,, \quad
Y_{\mathrm{cm}} = \dfrac{\int y\,dm}{m_{\mathrm{s}}}\,, \quad
Z_{\mathrm{cm}} = \dfrac{\int z\,dm}{m_{\mathrm{s}}}\,.
\end{align}
If the spring is made of an integral number of turns (i.e., complete loops),
symmetry implies that $X_{\mathrm{cm}} = Y_{\mathrm{cm}} = 0$. However, we must
calculate the last integral in equation~\eqref{Eq:CM} to find the vertical
position of the center of mass. This can be done easily if the element of mass
$dm$ is written as
\begin{align*}
dm = \lambda\,ds = \dfrac{m_{\mathrm{s}}}{S}\,\sqrt{r^2 + c^2}\,d\theta
   = \dfrac{m_{\mathrm{s}}}{2 \pi N}\,d\theta\,.
\end{align*}
Therefore
\begin{align}\label{Eq:Zcm_1}
Z_{\mathrm{cm}} = \dfrac{\int z\,dm}{m_{\mathrm{s}}}
 = \dfrac{1}{2 \pi N} \int_0^{2 \pi N}
\left[ c\,\theta + \dfrac{m_{\mathrm{s}}\,g\,(2\,L_0 -
c\,\theta)\,c\,\theta}{2\,k\,L_0^2}
\right]\,d\theta
 = \dfrac{L_0}{2} + \dfrac{m_{\mathrm{s}}\,g}{3\,k}\,.
\end{align}

\begin{figure}
\begin{center}
\subfigure[Horizontal spring.]{
\includegraphics[scale=0.8]{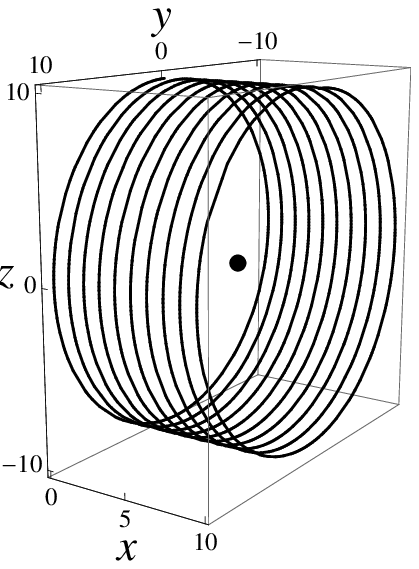}}
\hskip 3.0cm
\subfigure[Suspended spring.]{
\includegraphics[scale=0.8]{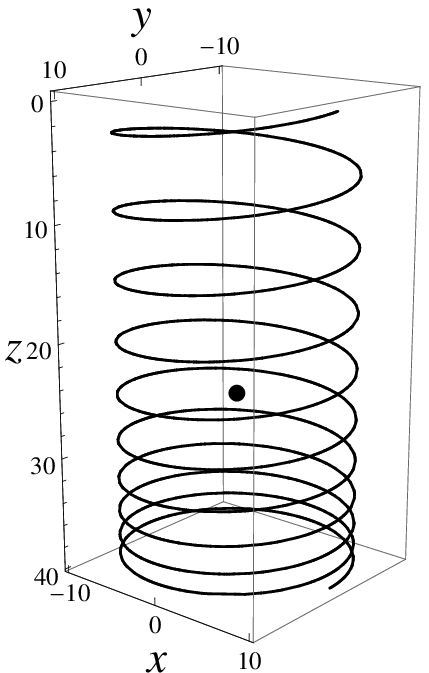}}
\caption{\label{Fig:helixes} Two springs with center of mass.}
\end{center}
\end{figure}

Galloni~\cite{Galloni:1076} and Mak~\cite{Mak:994}, in separate studies of the
static and dynamic effects of the mass of a soft spring, found that the total
elongation due to gravity of a vertically suspended spring is given by $\Delta L
= m_{\mathrm{s}}\,g / 2\,k$. Using this result, equation~\eqref{Eq:Zcm_1}
becomes
\begin{align}\label{Eq:Zcm_2}
Z_{\mathrm{cm}} = \dfrac{L_0}{2} + \dfrac{2}{3}\,\Delta L\,.
\end{align}
In other words, the center of mass of the suspended spring is displaced from its
position when not suspended by two-thirds of its elongation due to gravity (see
Figure~\ref{Fig:helixes}).

\section{Concluding remarks}
This particular problem is an excellent pedagogic tool for beginning students in
mathematics and physics. In particular, it shows the usefulness of a variety of
calculus concepts applied to a physical problem to which all students can
relate.

It is instructive to compare our results with some other models. For example,
the results of our calculations are in agreement with experimental observations
by Toepker~\cite{Toepker:16}. Equation~\eqref{Eq:Zcm_2} is also in accord with
equations obtained by Newburgh and Andes~\cite{Newburgh:586}, and
Ruby~\cite{Ruby:140,Ruby:324}, who used a hybrid calculus/non-calculus approach.
Equation~\eqref{Eq:Zcm_2} also agrees with the center of mass equation of a
suspended Slinky obtained by Hosken~\cite{Hosken:327}, in his revision of
French's work~\cite{French:244}. It is important to note that there is a
difference from their center of mass equation to ours by a factor of $1/2$. This
is easily explained by the fact that these authors did not incorporate the $1/2$
correction to the total spring mass explained in the works of
Galloni~\cite{Galloni:1076} and Mak~\cite{Mak:994}. On the other hand, our
approach does not concur with the model proposed by Sawicki~\cite{Sawicki:276}.
He found that the center of mass should be located at a distance $L_0/2 +
3\Delta L/4$ measured from the point of suspension.


\end{document}